\documentclass[conference]{IEEEtran}
\usepackage{amssymb}
\usepackage{amsthm}
\usepackage{multirow}
\usepackage{inputenc}
\usepackage{enumerate} 
\usepackage{textcomp}
\usepackage{listings}
\usepackage{array}
\usepackage[switch,mathlines]{lineno}
\usepackage{soul}
\usepackage{xfrac}
\usepackage{graphicx}
\usepackage{cite}
\usepackage[cmex10]{amsmath}
\usepackage{algorithm}
\usepackage{xcolor}
\usepackage{colortbl}
\usepackage{makecell}
\usepackage{cancel}

\usepackage{algorithm,algpseudocode}% http://ctan.org/pkg/{algorithms,algorithmx}
\algnewcommand{\Inputs}[1]{%
  \State \textbf{Inputs:}
  \Statex \hspace*{\algorithmicindent}\parbox[t]{.8\linewidth}{\raggedright #1}
}
\algnewcommand{\Initialize}[1]{%
  \State \textbf{Initialize:}
  \Statex \hspace*{\algorithmicindent}\parbox[t]{.8\linewidth}{\raggedright #1}
}

\makeatletter
\def\footnoterule{\kern-3\p@
  \hrule \@width 3.3in \kern 2.6\p@} % the \hrule is .4pt high
\makeatother

\usepackage{algpseudocode}

\usepackage{stackengine}

\usepackage{algpseudocode}
\usepackage{CJK}
\usepackage[cmex10]{amsmath}
\usepackage{bm}
\usepackage{color}
\usepackage{amsmath,amsthm,amssymb,amsfonts}
\usepackage{flushend}
\usepackage{float}

\usepackage{arydshln} %to include dashed lines in arrays
\usepackage{hyperref}
\hypersetup{
     colorlinks   = true,
     linkcolor    = blue,
     citecolor    = red,
     urlcolor     = blue
}
\makeatletter
\newcommand*{\transpose}{%
  {\mathpalette\@transpose{}}%
}
\newcommand*{\@transpose}[2]{%
  \raisebox{\depth}{$\m@th#1\intercal$}%
}
\makeatother

\usepackage{soul}
\usepackage{tikz}
\usepackage{booktabs}% http://ctan.org/pkg/booktabs
\usepackage{tabularx}% http://ctan.org/pkg/tabularx

\newcommand*\diff{\mathop{}\!\mathrm{d}}

\usepackage[caption=false,font=footnotesize]{subfig}
\ifCLASSINFOpdf
\else
\fi

\begin{document}
\renewcommand{\ttdefault}{cmtt}
\bstctlcite{IEEEexample:BSTcontrol}

\IEEEoverridecommandlockouts
% The preceding line is only needed to identify funding in the first footnote. If that is unneeded, please comment it out.

\def\BibTeX{{\rm B\kern-.05em{\sc i\kern-.025em b}\kern-.08em
    T\kern-.1667em\lower.7ex\hbox{E}\kern-.125emX}}

\title{Fast Critical Clearing Time Calculation for Power Systems with Synchronous and Asynchronous Generation
\thanks{This work was supported by the Natural Science Foundation of Jiangsu Province under Grant BK20230851. Corresponding author: Yijun Xu}
}       
% <-this % stops a space

% <-this % stops a space

% \author{
% {Xuezao Wang,~\IEEEmembership{Student Member},
% Yijun Xu,~\IEEEmembership{Senior Member}, 
% Wei Gu,~\IEEEmembership{Senior Member},
% Kai Liu,
% Shuai Lu,\\
% Mert Korkali,~\IEEEmembership{Senior Member}, and
% Lamine Mili, \IEEEmembership{Life Fellow}
% }
% %~\IEEEmembership{Student Member},
%        \thanks{X. Wang, Y. Xu, W. Gu, K. Liu, S. Lu are with the School of Electrical Engineering, Southeast University, Nanjing, Jiangsu 210096, China, (e-mail: \texttt{\{xuezaowang, yijunxu, wgu, kailiu02, shuai\_lu\}@seu.edu.cn}).} \thanks{M.~Korkali is with the University of Missouri, Department of Electrical Engineering and Computer Science, Columbia, MO 65211 USA (e-mail: \texttt{korkalim@missouri.edu}).}

% \thanks{L. Mili is with the Bradley Department of Electrical and Computer Engineering, Virginia Tech, Northern Virginia Center, Falls Church, VA 22043 USA (e-mail: \texttt{lmili@vt.edu}).}
% \thanks{This work was supported by the Natural Science Foundation of Jiangsu Province under Grant BK20230851.}
% %Document released as LLNL-JRNL-820867.\ %\emph{(Corresponding author: Zongsheng Zheng.)
% %}
% }

\author{
\IEEEauthorblockN{Xuezao Wang, Yijun Xu,\\
Wei Gu, Kai Liu,\\
and Shuai Lu}
\IEEEauthorblockA{\textit{School of Electrical Engineering} \\
\textit{Southeast University}\\ 
Nanjing, Jiangsu 210096, China \\
\{xuezaowang, yijunxu, wgu,\\ kailiu02, shuai\_lu\}@seu.edu.cn}
\and
\IEEEauthorblockN{Mert Korkali}
\IEEEauthorblockA{\textit{Department of Electrical Engineering}\\ 
\textit{and Computer Science}\\ 
\textit{University of Missouri} \\
Columbia, MO 65211 USA \\
korkalim@missouri.edu}
\and
\IEEEauthorblockN{Lamine Mili}
\IEEEauthorblockA{\textit{Bradley Department of Electrical}\\ 
\textit{and Computer Engineering}\\ 
\textit{Virginia Tech}\\
\textit{Northern Virginia Center} \\
Falls Church, VA 22043 USA \\
lmili@vt.edu}
}

% The paper headers
%\markboth{IEEE TRANSACTIONS ON CIRCUITS AND SYSTEMS—II: EXPRESS BRIEFS}%
%{Wang \MakeLowercase{\textit{\textit{et al.}}}: Fast CCT Calculation for Power Systems with Synchronous and Asynchronous Generations}
\maketitle

\begin{abstract}
%With the integration of renewable energy, modern power systems have become more unstable.
The increasing penetration of renewables is replacing traditional synchronous generation in modern power systems with low-inertia asynchronous converter-interfaced generators (CIGs). This penetration threatens the dynamic stability of the modern power system. To assess the latter, we resort to the critical clearing time (CCT) as a stability index, which is typically computed through a large number of time-domain simulations. This is especially true for CIG-embedded power systems, where the complexity of the model is further increased. To alleviate the computing burden, we developed a trajectory sensitivity-based method for assessing the CCT in power systems with synchronous and asynchronous generators. This allows us to obtain the CCT cost-effectively. The simulation results reveal the excellent performance of the proposed method.
\end{abstract}


\IEEEpeerreviewmaketitle
% \vspace{-0.2cm}

\section{Introduction}
%\subsection{Motivations}
% \IEEEPARstart{N}{owadays}, 
Today, the penetration of renewables in modern power grids has increased significantly. %\cite{xiong2019nonsmooth,miao2022fifthorder}.
This penetration transforms the power system from synchronous to asynchronous generation, where the converter-interfaced generators (CIGs) play the key role \cite{yin2021stability}. 
However, since CIGs typically have much less inertia than traditional synchronous generators\cite{liu2024bayesian}, 
their penetration into modern power systems challenges their dynamic stability. 
The critical clearing time (CCT) is a widely adopted tool for properly assessing the dynamic stability of a power system since it indicates its ability to tolerate a fault over a period of time.   However, its computing process becomes time-consuming for large-scale systems since it is typically based on repeated evaluations of many time-domain simulations that do not meet the practical demand. This is especially true for the modern power system, which simultaneously has synchronous and asynchronous generators that significantly increase its model complexity. Therefore, we choose it as the focus of this paper to develop a cost-effective CCT evaluation tool for modern power systems. 

%\subsection{Literature Review}
Previously, many researchers have investigated different strategies for calculating the CCT. Among them, the time domain simulation is the most widely used approach for its flexibility in different types of models \cite{kundur2022power}. Then, to further improve its computing efficiency, Yorino \emph{et al.} proposed a trapezoidal strategy to accelerate the numerical integration in calculating the CCT \cite{yorino2010new}.  However, when randomness in the system is considered \cite{6153044}, its computing burden is inevitably heavy. Early investigations have promoted simulation-free energy-function-based approaches for accelerated assessment of CCT \cite{Athay1979A}. 
For example, Thanh \emph{et al.} suggested a strategy based on the boundary of fault dynamics and the expansion of the Lyapunov function family. By solving a convex optimization problem, they obtained the lower bound of the CCT \cite{vu2016toward}.
Wu \emph{et al.} further proposed a method for estimating the CCT that relies on the structural analysis of the transient energy of the power system \cite{wu2019structural}. 
Although computationally efficient, these energy function-based methods face challenges in accurately determining the critical energy boundary \cite{vu2015lyapunov} and are limited to the power system with traditional synchronous generators. When it comes to the modern power system, where both synchronous and asynchronous generators are connected, the corresponding analytical derivations do not hold anymore. 

In summary, traditional time-domain simulation is computationally expensive due to repeated system trajectory evaluations, and energy function-based methods are primarily applicable to systems dominated by synchronous generators. There is an urgent need to propose a solution with greater computational efficiency and broader applicability.
Therefore, we develop a cost-effective method that achieves a fast CCT evaluation capability and applies to the CIG-embedded power system model. Specifically,  the following contributions are made:
\begin{itemize} 
\item By considering both synchronous and asynchronous grid-following (GFL) generators, we analytically derive the CCT solution for the modern power system using the trajectory sensitivity-based method. This allows us to accurately evaluate the CCT with only a few evaluations. 

\item The trajectory sensitivity method further enables us to approximate the stability of the CIG-embedded power system. The simulation results reveal that increasing the penetration of the GFL-based converters, in general, decreases the stability of the power system.
\end{itemize}

%\subsection{Paper Organization}
%The remainder of this paper is organized as follows. Section II describes the formulation of the problem. Section III elaborates on the proposed method for calculating the CCT for the CIG-embedded power system. Case studies are conducted in Section IV, followed by the conclusions in Section V.

\vspace{-0.2cm}
\section{Preliminaries}
This section describes a dynamic model of a power system with 
synchronous and asynchronous generators. Then, CCT as a stability index is introduced.

\subsection{The Dynamic Model for Modern Power Systems}
\subsubsection{Synchronous Generators}
In the traditional power system, the classical model for the synchronous generators is modeled as \cite{sauer1998power}
\begin{equation}
\label{Swing1}
\frac{\diff\delta}{\diff t}=\omega-\omega_0,
\end{equation}
\begin{equation}
\label{Swing2}
\frac{2H}{\omega_0}\frac{\diff\omega}{\diff t}=P_M-P_e-D(\omega-\omega_0).
\end{equation}
\noindent
Here, $\omega_0$ is the rated rotor angular frequency. Mechanical and electrical power inputs are denoted by $P_M$ and $P_e$, respectively. The damping ratio is $D$, and the inertia time constant is denoted by $H$. The state variables include the rotor angle and angular frequency, i.e., $\delta$ and $\omega$, respectively. 

\subsubsection{Asynchronous Generators}
Nowadays, the CIGs are gradually replacing traditional synchronous generators. The GFL converter allows for integrating renewable sources, such as solar and wind power, into the power grid as asynchronous generators \cite{liu2024bayesian}. 
Let us consider the dynamic power system model proposed by \cite{guo2021estimation,liu2024bayesian}. Let us take the reference value of the output active power $P_v$ as $P_{vs}$. The time constants for the virtual inertia control and CIG are $T_v$ and $T_p$, respectively. The intermediate variable is $x_v$.
The virtual inertia time constant is $H_v$. 
Thus,  the GFL power converter can be depicted as
\begin{equation}
\label{GFL1}
\frac{\diff x_v}{\diff t}=\frac1{T_v}(\omega_{P}-x_v),
\end{equation}
\begin{equation}
\label{GFL2}
\frac{\diff P_v}{\diff t}=\frac1{T_p}\left(P_{vs}-2H_v\frac{\diff x_v}{\diff t}-P_v\right).
\end{equation}

To guarantee that the output current of GFL converters is in phase with the grid voltage, the phase and frequency of the grid voltage are tracked by a phase-locked loop (PLL). Ideally, the output angle $\theta_{P}$ of the PLL should align with the system voltage angle, $\theta$. The proportional and integral gains of the PLL controller are denoted by the coefficients $K_p$ and $K_i$, respectively.  The intermediate state variable is $x_{P}$. The q-axis voltage is denoted as $v_q$. The difference between the rated rotor angular frequency, $\omega_0$, and the measured frequency at the connecting point is $\omega_{P}$. Thus, we have the PLL model as
\begin{equation}
\label{PLL1}
\frac{\diff\theta_{P}}{\diff t}=\omega_{P}+\omega_0,
\end{equation}
\begin{equation}
\label{PLL2}
\frac{\diff x_{P}}{\diff t}=v_q=V\sin(\theta-\theta_{P}),
\end{equation}
where $\omega_P=K_pv_q+K_ix_{P}$. 

\subsection{CCT as the Evaluation Metric for System Stability}
The capacity of power systems to maintain a stable operational condition following a major disturbance, such as a fault and load shedding, is known as \emph{transient stability}. Some critical indices are typically adopted to assess it. Among them, the CCT is widely adopted, as it quantifies the maximum duration of a fault the system can endure while maintaining stability. If the fault is cleared within the CCT, the system can regain stability; if the fault persists beyond the CCT, the system will lose its stability. Therefore, we select the CCT as an important metric for evaluating the transient stability of the modern power system.

\section{The Proposed Method for Calculating CCT for Synchronous and Asynchronous Generation}
In this section, we first derive the trajectory sensitivity of the CIG-embedded power system to the fault-clearing time and propose a novel method for accurately estimating the CCT of faults in power systems with CIGs.

\subsection{The CIG-embedded  Power System Trajectory Sensitivity to Fault Clearing Time}
The trajectory sensitivity theory was initiated by Hiskens \emph{et al.} \cite{hiskens2000trajectory} to study the variations of system state variables concerning small changes in initial conditions or parameters. Using it, importance ranking, uncertainty approximation, and inverse problems can be carried out in power systems. 

To this end, modern power systems with synchronous and asynchronous generation are represented by a set of differential and algebraic equations expressed as
\begin{equation}
\label{DAE1}
\dot{\bm{x}}=P(\bm{x},\bm{y}),\quad 0=S(\bm{x},\bm{y}),
\end{equation}
\noindent
where the function $P(\cdot)$ describes the dynamic behaviors of the system while the function $S(\cdot)$ represents the algebraic power flow equations. $\bm{x}$ is the vector of state variables with $\bm{x}=[\delta, \omega, x_v, P_v, \theta_P, x_P]$; $\bm{y}$ is the vector of algebraic variables with $\bm{y}=[P_e, \dot{I}, V, \theta]$.
Accordingly, the solution for \eqref{DAE1} is given by
\begin{equation}
\label{solu-DAE}
\bm{x}(t)=w({\bm{x}_0},t),\quad 
\bm{y}(t)=u({\bm{x}_0},t),
\end{equation}
where $w(\cdot)$ and $u(\cdot)$ represent the system trajectory given initial conditions, ${\bm{x}_0}$ and ${\bm{y}_0}$.

To analyze fault clearing time, $T_{cl}$, in the CCT analysis, let us further define two time periods, namely, during the fault and after the fault, where the system is governed by
\begin{equation}
    \begin{aligned}&\bm{\dot{x}}=P_{1}(\bm{x},\bm{y}),\quad 0=S_{1}(\bm{x},\bm{y}),\\&\bm{x}(t_{1})=\bm{x}_{1},\quad \bm{y}(t_{1})=\bm{y}_{1},\end{aligned}\quad t\in[t_{1},t_{cl}],\label{d-dae}
\end{equation}
    \\
and by 
\begin{equation} 
  \begin{aligned}&\bm{\dot{x}}=P_{2}(\bm{x},\bm{y}),\quad 0=S_{2}(\bm{x},\bm{y}),\\&\bm{x}(t_{cl})=\bm{x}_{cl},\quad \bm{y}(t_{cl})=\bm{y}_{cl},\end{aligned}\quad t\in[t_{cl},\infty).\label{a-dae}
\end{equation}

\noindent
%where $t_0$ represents the time at fault starting, $T_{cl}$ represents the time at fault clearing. 
Here, $\bm{x}_1$ and $\bm{y}_1$ represent the system state at the inception of a fault, ${\bm{x}_{cl}}$ and ${\bm{y}_{cl}}$ represent the system state at fault clearing, respectively, and $P_1(\cdot)$ and $S_1(\cdot)$ and $P_2(\cdot)$ and $S_2(\cdot)$ represent the functions of the system during and after the fault, respectively. The fault occurs at time $t_1$ and is cleared at time $t_{cl}$. Therefore, the fault clearing time $T_{cl}$ can be expressed as $T_{cl}=t_{cl}-t_1$. If $T_{cl}$ equals the CCT ($T_{cr}$), the system remains stable for $t<t_1$ and becomes unstable for $t\geq{t_1}$. 

By applying Taylor series expansion on the solutions, $w$ and $u$, of trajectory sensitivities about $T_{cl}$, we get
\begin{equation}
\begin{aligned}
\Delta \bm{x}(t)& =\Delta w({\bm{x}_{cl}},t)\\&=  
 \frac{\partial w({\bm{x}_{cl}},t)}{\partial T_{cl}}\Delta T_{cl}+{\mathcal{E}}^{w}(\bm{x}_{cl},t,\Delta T_{cl}) ,\quad t\in[t_{cl},\infty),
\end{aligned}
\end{equation}
\begin{equation}
\begin{aligned}
\Delta \bm{y}(t)& =\Delta u({\bm{x}_{cl}},t)\\&=  
 \frac{\partial u({\bm{x}_{cl}},t)}{\partial T_{cl}}\Delta T_{cl}+{\mathcal{E}}^{u}(\bm{x}_{cl},t,\Delta T_{cl}),\quad t\in[t_{cl},\infty),
\end{aligned}
\end{equation}
where ${\mathcal{E}}^{w}$ and ${\mathcal{E}}^{u}$ are the higher order terms that can be ignored when $\Delta T_{cl}$ is small. It yields
\begin{equation}
\begin{aligned}
\Delta \bm{x}(t)\approx\frac{\partial w({\bm{x}_{cl}},t)}{\partial T_{cl}}\Delta T_{cl}\equiv W({\bm{x}_{cl}},t)\Delta T_{cl},
\end{aligned}
\end{equation}
\begin{equation}
\begin{aligned}
\Delta \bm{y}(t)\approx\frac{\partial u({\bm{x}_{cl}},t)}{\partial T_{cl}}\Delta T_{cl}\equiv U({\bm{x}_{cl}},t)\Delta T_{cl},
\end{aligned}
\end{equation}
where $W(\cdot)$ and $U(\cdot)$ are the trajectory sensitivity matrix for fault clearing time. Using \eqref{solu-DAE}, we have
\begin{equation}
\label{WU}
\begin{aligned}
W({\bm{x}_{cl}},t)=\frac{\partial w({\bm{x}_{cl}},t)}{\partial T_{cl}}=\frac{\partial \bm{x}(t)}{\partial T_{cl}},\\
U({\bm{x}_{cl}},t)=\frac{\partial u({\bm{x}_{cl}},t)}{\partial T_{cl}}=\frac{\partial \bm{y}(t)}{\partial T_{cl}}.
\end{aligned}
\end{equation}

So far, we have provided the definition of trajectory sensitivity. Since in the CCT, our main focus is the calculation of $T_{cr}$ for dynamic stability analysis, we will focus on the dynamic behavior of the system after the fault. So, we further compute the post-fault trajectory sensitivities.

By differentiating the post-fault system governed by \eqref{a-dae} with respect to $T_{cl}$, we get
\begin{equation}
\label{huan}
\begin{aligned}
\frac{\partial}{\partial t}\bigg(\frac{\partial \bm{x}(t)}{\partial T_{cl}}\bigg)=\frac{\partial P_{2}}{\partial \bm{x}} \frac{\partial \bm{x}(t)}{\partial T_{cl}}+\frac{\partial P_{2}}{\partial \bm{y}} \frac{\partial \bm{y}(t)}{\partial T_{cl}},\\0=\frac{\partial S_{2}}{\partial \bm{x}} \frac{\partial \bm{x}(t)}{\partial T_{cl}}+\frac{\partial S_{2}}{\partial \bm{y}} \frac{\partial \bm{y}(t)}{\partial T_{cl}}.
\end{aligned}
\end{equation}
Substituting \eqref{WU} into \eqref{huan}, we have
\begin{equation}
\begin{aligned}
\frac{\partial W({\bm{x}_{cl}},t)}{\partial t}=\frac{\partial P_{2}}{\partial \bm{x}} W({\bm{x}_{cl}},t)+\frac{\partial P_{2}}{\partial \bm{y}} U({\bm{x}_{cl}},t),\\0=\frac{\partial S_{2}}{\partial \bm{x}} W({\bm{x}_{cl}},t)+\frac{\partial S_{2}}{\partial \bm{y}} U({\bm{x}_{cl}},t).
\label{tttsa}
\end{aligned}
\end{equation}
The solution to \eqref{tttsa} is the CIG-embedded power system trajectory sensitivity for fault clearing time. 
To solve it, we still need to compute its initial conditions, $W({\bm{x}_{cl}},t_{cl}^+)$ and $U({\bm{x}_{cl}},t_{cl}^+)$.  More specifically, at $t_{cl}$, the state variables remain continuous once the fault is cleared. Thus, we have
\begin{equation}
    \bm{x}(t_{_{cl}}^+)=\bm{x}(t_{_{cl}}^-).
\label{xcl}
\end{equation}
Since $t_{cl}$ does not depend on the system parameters and variables, the trajectory sensitivity remains continuous, yielding
\begin{equation}
W(\bm{x}_{cl},t_{cl}^+)=W(\bm{x}_{cl},t_{cl}^-).
\label{W}
\end{equation}
By inserting \eqref{W} into post-fault system, we get
\begin{equation}
    0=\frac{\partial S_{2}}{\partial \bm{x}} W({\bm{x}_{cl}},t_{cl}^+)+\frac{\partial S_{2}}{\partial \bm{y}} U({\bm{x}_{cl}},t_{cl}^+).
\end{equation}
Assuming $\frac{\partial S_{2}}{\partial \bm{y}}$ is nonsingular, we have
\begin{equation}
    U({\bm{x}_{cl}},t_{cl}^+)=-\bigg(\frac{\partial S_{2}}{\partial \bm{y}}\bigg)^{-1}\frac{\partial S_{2}}{\partial \bm{x}} W({\bm{x}_{cl}},t_{cl}^+).
    \label{U}
\end{equation}
Now,  we obtain the trajectory sensitivity of the post-fault CIG-embedded power system equations to $T_{cl}$.

\subsection{CCT Calculation Considering Synchronous and Asynchronous Generation}
Here, our work investigates the CCT of modern power systems with synchronous and asynchronous generation. To utilize \eqref{U} for fast calculation of the CCT, $T_{cr}$, we further resort to an $\ell_2$-norm-based trajectory sensitivity index, $SN$, defined as $SN=(\sum_{i=1}^nx_i^2)^{\frac12}$, where $\bm{x}=(x_{1},x_{2},\dots,x_{n})^{\transpose}$ \cite{nguyen2002sensitivity}.

For the $m$ synchronous generators in the system, the $\ell_2$-norm $SN$ is computed as 
\begin{equation}
\label{SNsg}
SN_{SG}=\Bigg(\sum_{i=1}^{m}\bigg(\frac{\partial\delta_{i}}{\partial T_{cl}}-\frac{\partial\delta_{j}}{\partial T_{cl}}\bigg)^{2}+\bigg(\frac{\partial\Delta \omega_{i}}{\partial T_{cl}}\bigg)^{2}\Bigg)^{\frac12},
\end{equation}
\noindent
where the $j$th machine is the reference synchronous generator. Similarly, for the $n$ asynchronous GFL generators in the system, we have
\begin{equation}
\label{SNgfl}
\begin{split}
SN_{AG}=\Bigg(\sum_{i=1}^{n}\bigg(\frac{\partial\Delta x_{v_i}}{\partial T_{cl}}\bigg)^{2}+\bigg(\frac{\partial\Delta x_{P_i}}{\partial T_{cl}}\bigg)^{2}+\\
\bigg(\frac{\partial\Delta P_{v_i}}{\partial T_{cl}}\bigg)^{2}+\bigg(\frac{\partial\theta_{P_i}}{\partial T_{cl}}-\frac{\partial\theta_{P_k}}{\partial T_{cl}}\bigg)^{2}\Bigg)^{\frac12},
\end{split}
\end{equation}
where the $k$th machine is the reference asynchronous GFL generator. For a stable system, this norm will gradually approach zero. However, as the fault clearing time $T_{cl}$ approaches the CCT, $T_{cr}$, $SN$ throughout the transition process remains high.

Here, let us define the peak value of $SN$ as $m(SN)$. Using it, we define the sensitivity index, $\lambda$, for the system as
\begin{equation}
\lambda=\frac1{m(SN)}.
\end{equation}
Obviously, near the stability margin,  $m(SN)$ will be large and $\lambda$ will approach zero. On the $T_{cl}$-$\lambda$ coordinate plane, $T_{cl}$ and $\lambda$ form an approximate linear relationship. Therefore, to calculate the CCT of a system, we can utilize this linear relationship for interpolation. The details of the proposed method are described in Algorithm~\ref{CCT1}.

\begin{algorithm}[!htbp]
\caption{CCT Calculation Considering Synchronous and Asynchronous Generators Based on the Proposed Method}
\label{CCT1}
\begin{algorithmic}[1]
\State For each generator type (synchronous and asynchronous) in the system, compute the values of $m(SN)$ under two different fault clearing times $T_{cl1}$ and $T_{cl2}$, using \eqref{SNsg} and \eqref{SNgfl}. Then, calculate the corresponding index $\lambda$ values, denoted as $\lambda_1$ and $\lambda_2$;
\State In the ($T_{cl}$, $\lambda$) coordinate plane, mark the points ($T_{cl1}$, $\lambda_{1}$) and ($T_{cl2}$, $\lambda_{2}$);
\State Using the extrapolation method, determine the intersection of the line passing through these two points with the $T_{cl}$-axis. This intersection represents the CCT for each type of generation.
\State Select the smaller of the two results as the system's CCT for the fault.
\end{algorithmic}
\end{algorithm}

%\vspace{0.4cm}
The proposed method remains effective for systems with moderate penetration of the GFL converter. However, since GFL generators do not inherently provide inertia, higher penetration levels may lead to faster frequency deviations, affecting the accuracy of the trajectory sensitivity. Furthermore, PLL dynamics can introduce small deviations in the CCT estimation.   Future work may investigate higher-order trajectory sensitivity to improve its applicability.

\section{Case Studies}
 In this section, we apply the proposed method to a CIG-embedded power system to test its performance. In addition, we analyze the impact of the CCT caused by the GFL converters on a modified IEEE $39$-bus power system.
\vspace{-0.2cm}
\subsection{Calculation of CCT Using The Proposed Method}
\subsubsection{Test System Setup}
Let us first modify the test system by replacing the traditional synchronous generators \#36 and \#37 with asynchronous GFL generators.  The simulation time step is set to $0.01$ s, with a total simulation duration of 15 seconds. The system fault setup is as follows: at 5 seconds, a three-phase short circuit occurs at Bus $2$. This fault is cleared by disconnecting Line $2$-$3$.

The key parameters for the CIG model are listed in Table~\ref{tab0}.
%\textcolor{red}{These parameters are selected based on practical experience and remain valid for typical GFL converter applications. However, as the penetration level of the CIG increases, additional stability assessments should be considered.}
These parameters are selected based on practical experience and remain valid for typical GFL converter applications. The values in curly brackets in Table~\ref{tab0} correspond to the parameters of the two GFL generators with different power capacities. However, as the penetration level of the CIG increases, additional stability assessments should be considered.

%\begin{figure}[htbp]
%\centering
%\includegraphics[scale=0.28]{figure/cig39.png}
%\caption{Plot for the topology of the modified IEEE $39$-bus system.} 
%\label{39bus}
%\end {figure}
\vspace{-0.5cm}
\begin{table}[htpb!]
\centering
\caption{Parameters of the CIGs}
\begin{tabular}{cccc}
\hline
\textbf{Parameters} & \textbf{Value}   & \textbf{Parameters} & \textbf{Value}   \\ \hline
${P_{vs}}$        & $\{560,540\}$ & $H_{v}$         & $\{7.5,5\}$   \\
${T_{v}}$         & $\{2,2\}$     & $K_{p}$         & $\{10,10\}$   \\
${T_{p}}$         & $\{1,1\}$     & $K_{i}$         & $\{100,100\}$ \\ \hline
\end{tabular}
\label{tab0}
\end{table}

\subsubsection{Demonstration of the Proposed Method}
\begin{figure}
    \centering
    \includegraphics[scale=0.6]{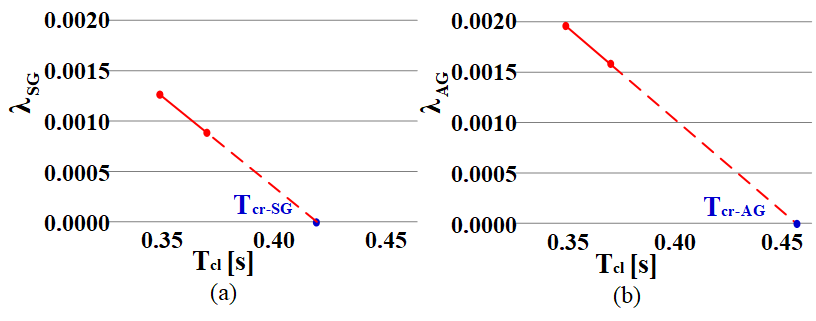}
    \vspace{-0.5cm}
    \caption{CCT calculation of CIG-embedded power system using the proposed method for (a) synchronous generators and (b) asynchronous generators.}
    \label{cal_tcl}
\end{figure}

Let us demonstrate CCT computing using our proposed method. For example, we randomly choose $0.35$ s and $0.37$ s for a demonstration to compute $m(SN)$ and $\lambda$.  As shown in Table~\ref{tab1}, we use the detailed values of these two points, $T_{cl}$ and $\lambda$ at $0.35$ s and $0.37$ s, to further calculate the CCTs for both synchronous and asynchronous generators, respectively, from extrapolation. Finally, the CCT for synchronous generators is computed as $0.4171$ s, and the CCT for asynchronous generators is computed as $0.4565$ s. A lower value of $0.4171$ s is selected as the system's CCT. The simulation results are in Fig.~\ref{cal_tcl}.  
\vspace{-0.5cm}
\begin{table}[htbp!]
\centering
\caption{Different values of $T_{cl}$ corresponding to $SN$ and $\lambda$}
\begin{tabular}{ccccc}
\hline
${T_{cl}}$ & ${m(SN_{SG})}$ & ${\lambda_{SG}}$ & ${m(SN_{AG})}$ & ${\lambda_{AG}}$      \\ \hline
$0.35$ s & $800.5685$  & $0.0012491$  & $504.1394$ & $0.0019836$ \\
$0.37$ s & $1{,}141.4653$ & $0.00087607$ & $619.2539$ & $0.0016148$ \\ \hline
\end{tabular}
\label{tab1}
\end{table}

\subsubsection{Validation of the Proposed Method}
\begin{figure}
    \centering
    \includegraphics[scale=0.48]{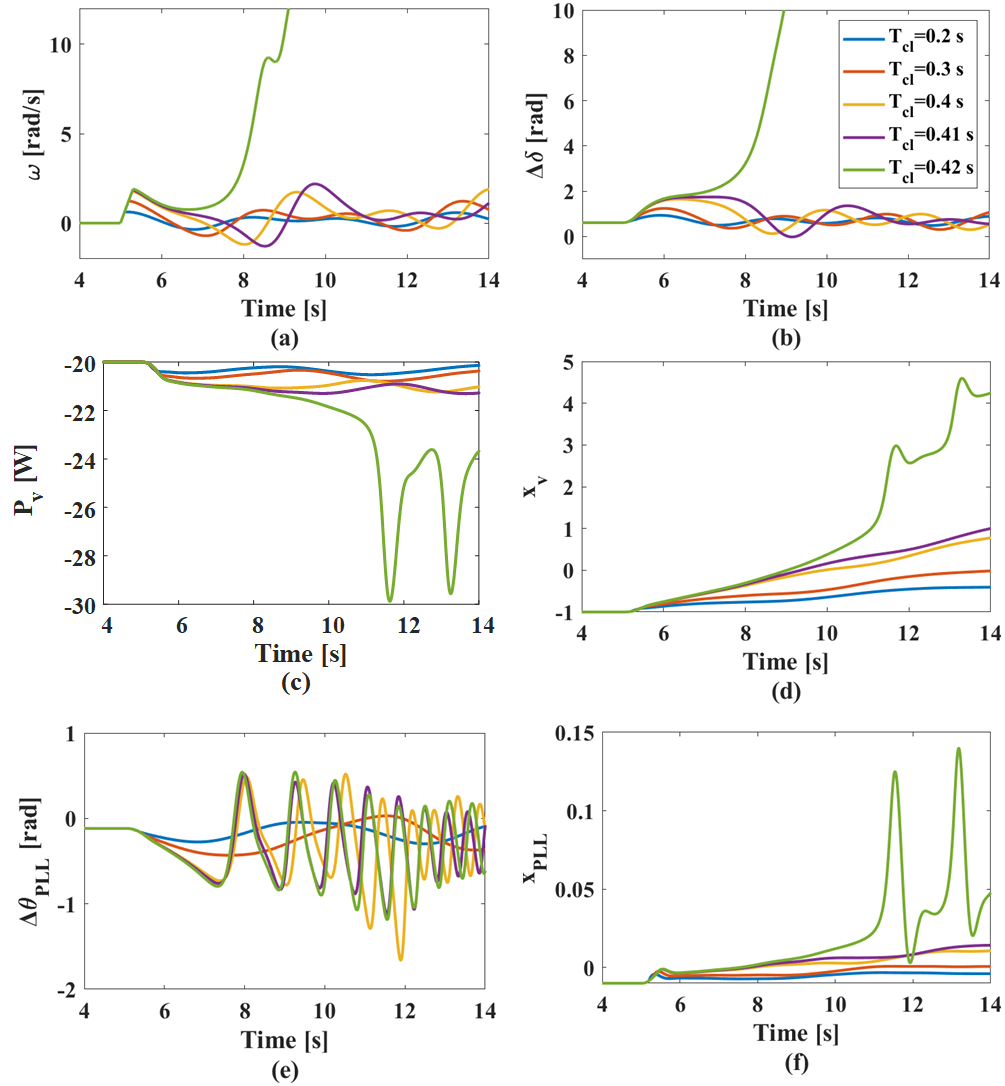}
    \vspace{-0.3cm}
    \caption{Simulation of state variables of CIG-embedded power system on the different values of $T_{cl}$ with its (a) and (b) for synchronous generators and (c)-(f) for asynchronous generators.}
    \label{gfl_tcl}
\end{figure}

Next, we validate the proposed method with the traditional time-domain simulation (TDS) method. We first empirically provide a possible range for the CCT. Then, using TDS, the interval is progressively narrowed using the interval elimination principle, leading to an approximate solution for the CCT. It is important to note that the CCT obtained here is not a single value but a small interval within which the precise solution is located. As shown in Fig.~\ref{gfl_tcl},  the fault in the system is cleared at different time intervals. It can be observed that as long as the fault clearing time is less than $0.41$ s, the system remains stable. However, when $T_{cl}=0.42$ s, the system loses stability. Thus, we can determine that the CCT of the CIG-embedded system is between $0.41$ s and $0.42$ s. The proposed method yields a CCT of $0.4171$ s, falling within the acceptable range. This result further validates the proposed method’s rationale. 

\subsubsection{Computing Efficiency}
The computing time of the CCT using the proposed method and the traditional TDS method takes $1.38$ s and $10.93$ s, respectively. This is because the proposed method only requires computation of trajectory sensitivity twice for the fault-clearing time. Therefore, it significantly improves the computational speed compared to the traditional method that requires repeated evaluation of the samples to approach the CCT. 

\subsubsection{Test under Different Events}
We apply faults at different buses to verify the applicability of the proposed method. The simulation results are shown in Table~\ref{tab2}, which demonstrates that the proposed method to calculate the CCT provides accurate results under different events. 
\vspace{-0.3cm}
\begin{table}[htpb!]
\centering
    \caption{Estimation of CCT for Faults at Different Buses}
\begin{tabular}{cccc}
\hline
\textbf{Fault Bus} & \textbf{Tripped Line} & \textbf{CCT (TDS)} & \textbf{CCT (Proposed Method)} \\ \hline
$3$         & $3$-$18$     & $0.37$-$0.38$ s & $0.3715$ s   \\
$7$         & $7$-$8$      & $0.38$-$0.39$ s & $0.3895$ s   \\
$14$        & $14$-$15$    & $0.36$-$0.37$ s & $0.3772$ s   \\
$26$        & $26$-$28$    & $0.27$-$0.28$ s & $0.2725$ s   \\ \hline
\end{tabular}
\label{tab2}
\end{table}

\subsection{The Impact of CIGs on Power Systems considering CCT}
\begin{figure}[htbp!]
    \centering
    \includegraphics[scale=0.52]{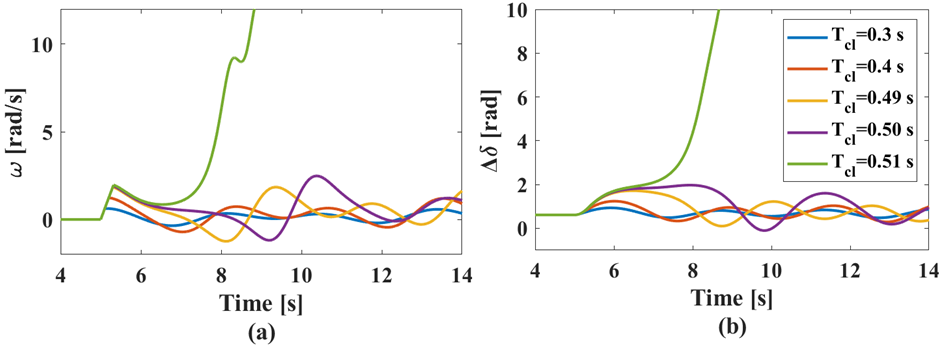}
    \vspace{-0.3cm}
    \caption{Simulation of state variables of traditional IEEE $39$-bus system on the different values of $T_{cl}$ with its (a) for $\omega$ and (b) for $\delta$.}
    \label{main_tcl}
\end{figure}

To better understand the impact that CIGs have on the stability of a power system, we further calculate the CCT on the IEEE $39$-bus system without asynchronous generators. Under the same fault, the simulation results are shown in Fig.~\ref{main_tcl}. The associated CCT for the fault is found to be between $0.50$-$0.51$ s using the TDS method, which is longer than the CCT of the power system with two asynchronous GFL generators discussed earlier. The CCT calculated by the proposed method is $0.5062$ s, which means that the proposed method obtains similar computing results.  

These simulation results indicate that CIG integration has reduced the stability of modern power systems, highlighting the urgent need to enhance their transient stability.
While, in specific scenarios, GFL converters may exhibit improved transient response due to their fast control dynamics, their increasing penetration eventually leads to reduced system inertia and a decline in overall transient stability.

\vspace{-0.1cm}

\section{Conclusion}
In this paper, we propose a method for calculating the CCT of power systems with both synchronous and asynchronous generators based on trajectory sensitivity analysis. This method can significantly shorten the calculation time of CCT compared to the traditional TDS method while maintaining high accuracy.

%{\color{red} Beyond its theoretical contributions, our method has significant practical implications for power system operation and control. First, its ability to rapidly estimate CCT enables real-time transient stability assessment, allowing system operators to quickly identify critical conditions and implement preventive measures. Second, in power grids with high penetration of converter-interfaced generation, our method provides an efficient tool for analyzing transient stability under various operating conditions, aiding in the optimization of dispatch and control strategies.}

Beyond theoretical contributions, our method supports the assessment of transient stability in real time, helping operators identify critical conditions and take preventive actions. In addition, it improves the stability analysis in high-CIG power systems.

To further enhance this research, future work will explore the extension of our method to grid-forming converters, which exhibit different transient characteristics compared to GFL inverters. Furthermore, incorporating machine learning techniques could further improve the adaptability and precision of CCT estimation under varying system conditions. Another potential research direction is to consider uncertainty modeling in renewable generation, allowing a more robust assessment of transient stability in practical power system applications. These extensions will help refine and expand the applicability of our proposed method to future power systems.

%\vspace{26em}  % 插入一定高度的空白空间

% \section*{Acknowledgment}
% This work was supported, in part, by the U.S. National Science Foundation under Grant 1917308 and by the United States Department of Energy Office of Electricity Advanced Grid Modeling Program, and performed under the auspices of the U.S. Department of Energy by Lawrence Livermore National Laboratory under Contract DE-AC52-07NA27344. Document released as LLNL-PROC-XXXXXX.

% \ifCLASSOPTIONcaptionsoff
%   \newpage
% \fi
 \vspace{-0.1cm}
 \newcommand{\BIBdecl}{\setlength{\itemsep}{0.01 em}}
 \bibliographystyle{IEEEtran}
\bibliography{IEEEabrv,References.bib}

\end{document}